\documentclass[twocolumn,groupedaddress]{revtex4}
\usepackage{colordvi,epsfig,color,amsmath,amssymb}

\begin{document}

\def\be{\begin{equation}}
\def\ee{\end{equation}}
\def\bee{\begin{eqnarray}}
\def\eee{\end{eqnarray}}
\def\sech{\mbox{sech}}
\def\e{{\rm e}}
\def\d{{\rm d}}
\def\L{{\cal L}}
\def\U{{\cal U}}
\def\M{{\cal M}}
\def\T{{\cal T}}
\def\V{{\cal V}}
\def\R{{\cal R}}
\def\kb{k_{\rm B}}
\def\tw{t_{\rm w}}
\def\ts{t_{\rm s}}
\def\Tc{T_{\rm c}}
\def\gs{\gamma_{\rm s}}
\def\tm{tunneling model }
\def\TM{tunneling model }
\def\tilde{\widetilde}
\def\Deltac{\Delta_{0\rm c}}
\def\Deltamin{\Delta_{0\rm min}}
\def\Emin{E_{\rm min}}
\def\tauc{\tau_{\rm c}}
\def\tauac{\tau_{\rm AC}}
\def\tauw{\tau_{\rm w}}
\def\taumin{\tau_{\rm min}}
\def\taumax{\tau_{\rm max}}
\def\de{\delta\varepsilon / \varepsilon}
\def\pF{{\bf pF}}
\def\pFAC{{\bf pF}_{\rm AC}}
\def\halb{\mbox{$\frac{1}{2}$}}
\def\dreihalb{\mbox{$\frac{3}{2}$}}
\def\viertel{\mbox{$\frac{1}{4}$}}
\def\achtel{\mbox{$\frac{1}{8}$}}
\def\with{\quad\mbox{with}\quad}
\def\und{\quad\mbox{and}\quad}
\def\za{\sigma_z^{(1)}}
\def\zb{\sigma_z^{(2)}}
\def\ya{\sigma_y^{(1)}}
\def\yb{\sigma_y^{(2)}}
\def\xa{\sigma_x^{(1)}}
\def\xb{\sigma_x^{(2)}}
\def\spur#1{\mbox{Tr}\left\{ #1\right\}}
\def\erwart#1{\left\langle #1 \right\rangle}
\newcommand{\bbbone}{{\mathchoice {\rm 1\mskip -4mu l}{\rm 1\mskip -4mu l}{\rm 1\mskip -4.5mu l}{\rm 1\mskip -5mu l}}}

\title{Breakdown of perturbative weak coupling approaches for the biomolecular energy transfer}

\author{P. Nalbach and M. Thorwart}
\affiliation{Freiburg Institute for Advanced Studies (FRIAS), School of Soft Matter Research, Albert-Ludwigs-Universit\"at Freiburg, Albertstr. 19, 79104 Freiburg, Germany}

\date{\today}

\begin{abstract}
We show that the biomolecular exciton dynamics under the influence of slow polarization fluctuations in the solvent cannot be described by approaches which are perturbative in the system-bath coupling. For this, we compare results for the decoherence rate of the exciton dynamics of a resumed perturbation theory with numerically exact real-time path-integral results. We find up to one order in magnitude difference in the decoherence rate for realistically slow solvent environments even in the weak coupling regime, while both results coincide for fast environmental noise. This shows explicitely the nonperturbative influence of the bioenvironmental fluctuations and might render current perturbative approaches to biomolecular exciton transport questionable.
\end{abstract}


\maketitle

\section{Introduction}

Photosynthesis in biomolecular complexes is a fundamental but at the same time rather complex series of processes which converts photonic energy of sunlight into chemical energy which is stored by an organism for later use to drive cellular processes
\cite{Blankenship,Amerongen}. At the beginning of this chain, a photon is absorbed by one of the light-harvesting antenna structures (pigments) and an exciton is created. Its excitation energy is subsequently transferred via several intermediate pigments within the antenna system to the reaction center where it is converted into chemical energy. The transfer of this excitation energy within the antenna structure occurs by migration of electronic excited states from one molecular part to another and is a purely physical process which astonishingly occurs with almost unity quantum yield.

The theoretical formulation for energy transfer which is nowadays widely applied to photosynthetic systems was provided by Theodor F\"orster in the 1940s \cite{Forster}. It holds for the case when the pigments are spatially rather distant and thus weakly coupled
by incoherent (diffusive) exciton flow and also when they are very close and thus strongly coupled by coherent exciton motion \cite{Knox93}. The latter is often referred to as exciton coupling and the resulting exciton-coupled dimer can effectively be viewed as a supermolecule with delocalized electronic transitions where the molecular wave functions are given as superposition of the wave functions of the two monomeric pigments. Despite the somewhat different situations, it has been shown that the two cases are just two sides of the same coin \cite{Knox93,Blankenship,Amerongen}. The F\"orster energy transfer
mechanism is a nonradiative resonance transfer process based on the Coulomb dipole-dipole interaction of the excitons in the
pigments.

The photosynthetic excitons are by no means isolated quantum objects, but are embedded in host protein structure and in the surrounding polar solvent. It has been shown \cite{Gilmore05} that decoherence due to the coupling of the electric dipole moment of the exciton to the fluctuating electric dipole moments of the individual molecules of the solvent is not negligible. Decoherence and energy relaxation
in the energy transfer processes in photosynthetic antenna complexes has been studied theoretically in various variants
\cite{Gilmore05,Amerongen,Leegwater96,Vulto99,Jang02,Jang04,Mohseni08,Caruso09,Rebentrost09,Sarovar09,Flemming1,Gilmore06,Gilmore08,Olaya1,Olaya2,Flemming2,May04,Breuer02,Nitzan,BioTh2008}. Essentially all works rely on the most commonly used approach which considers the exciton-environment coupling as weak and treats it perturbatively~\cite{foot4} in order to derive some form of a time-local quantity for the time evolution of the system. This commonly results in or is equivalent to the famous Redfield master equation \cite{Amerongen,May04,Breuer02,Nitzan,Flemming1}. Two central assumptions are made: (i) the system-environment coupling is sufficiently weak in order that a second-order perturbative description should be valid, and (ii) the bath-induced fluctuations of the excitonic energy levels are sufficiently fast so that they can be treated in a Markov sense. Only then, time-local evolution equations result and the problem is sufficiently simple for further, often numerical, treatment. Put differently, it is assumed that the phonons which generate the fluctuations relax very quickly to their respective equilibrium state. The time scale thereby is given by the reorganization energy $E_R$ which is assumed to be much larger than all other energy scales in the model. The reorganization energy is related to the cut-off frequency $\omega_c$ of the bath spectral density $J(\omega)$ and denotes the maximal frequency, up to
which environmental modes exist. For the most important fluctuating electric dipole moments
in the protein-solvent environment, the spectral distribution of the bath modes has an Ohmic form
\cite{Gilmore05,Gilmore06,Gilmore08}, i.e., $J(\omega)=2\alpha\omega\exp(-\omega/\omega_c)$ with coupling constant $\alpha$,
which is related to the characteristic quantities of the solvent, such as the dipole moment, the frequency dependent dielectric constant and the Debye relaxation time~\cite{foot5}. This implies $E_R=2\alpha \hbar \omega_c$. In order that the perturbative description is valid, the intrinsic exciton dynamics (which is characterized by the energy gap $\Delta$) has to be much slower than the bath reorganization processes, i.e, $\Delta \ll \omega_c$. Only then and with the additional weak coupling condition $E_R\ll \omega_c$, all environmental influence can be treated to lowest order in the interaction and as effectively time local.

This, however, is typically not the case for the photosynthetic excitation energy transfer. Typical energy scales in biomolecules are $\Delta \simeq 0.2-2$ meV for resonant energy transfer between red and green chromophores or
$\Delta \simeq 46-100$ meV for the light-harvesting complex LH-II in the bacteria chlorophyll molecule in green sulphur bacteria
\cite{Gilmore06}, while the environmental energy scale are $\hbar \omega_c\simeq 2-8$ meV for water at room temperature. Moreover, for typical free chromophores in water at room temperature, $\alpha\simeq 0.1 -1$.
This shows that one typically has to deal with cases where $\Delta \lesssim \omega_c$, implying that any perturbative approach to energy transfer in these structures becomes questionable. This has also been investigated in great detail by Ishizaki and Fleming \cite{Flemming1,Flemming2}. They found that none of the variants of the perturbative approaches (Redfield equation in its full form, in the secular approximation and with the neglect of the imaginary part of the relaxation terms) can give a reliable picture \cite{Flemming1}. In order to overcome the deficiencies, a reduced hierarchy equation approach has been proposed~\cite{Flemming2}. At the same time numerical exact treatments were used~\cite{BioTh2008} to describe
the effect of slow environmental fluctuations.

One of the experimentally best studied systems is the Fenna-Matthews-Olson (FMO) protein~\cite{Bio3,Bio4} which serves in green sulfur bacteria such as {\it Prosthecochloris aestuarii} and {\it Chlorobium tepidum} as exciton conductor between the antenna complex and the reaction center. The FMO protein is a trimer where the three identical subunits consist of seven chromophoric sites which are fully characterized~\cite{BioCh2005}. Typical transport times for excitons through FMO are of the order of picoseconds. Recent experiments~\cite{BioEn2007} on the FMO complex show that the exciton transport exhibits quantum coherent oscillations up to $660$ femtoseconds. This raises the important question why the strong environmental fluctuations are insufficient to fully decohere the quantum transport. Simplifying the problem to a donor-acceptor model level it was shown~\cite{BioTh2008,Flemming2} that the slowness of the environment, namely $\omega_c\simeq\Delta$, sufficiently slows down decoherence for quantum oscillations to survive even for strong coupling to the environment. 

Having understood why quantum coherence survives rather strong couplings to environmental fluctuations raises the subsequent question whether photosynthesis benefits from quantum coherence either to improve efficiency or to allow regulation. Detailed descriptions of transport through FMO, however, requires to go beyond a donor-acceptor model rendering the problem much more complicated. Therefore allmost all approaches to study this question rely on weak coupling approximations~\cite{Mohseni08,Olaya1,Olaya2,Caruso09,Rebentrost09,Sarovar09,Flemming1}. Ishizaki and Fleming \cite{Flemming2,Fleming3} use a reduced hierarchy approach which they match for weak coupling to results from a Redfield approach and for strong coupling to results from F\"orster transfer description. The actual parameter range of interest is in between. 
We show in the present paper that even at weak coupling perturbative weak coupling approaches (like Redfield) break down and fail to describe the dynamics.

We shortly review the Hamiltonian for a single chromophore in the next section~\ref{sec2} before we once more introduce a donor-acceptor model in order to discuss the validity of perturbative approaches by comparing a lowest order perturbative approach with a numerical exact treatment. Using a resumed perturbative approach (RESPET)~\cite{RESPET1,RESPET2,RESPET3,RESPET4,RESPET5} we analytically derive in section~\ref{sec3} the time evolution of the population difference in our donor-acceptor model in lowest order in the system environment coupling. Thereby all non-Markovian corrections are determined in lowest order and they turn out to be negligible. In section~\ref{sec4} the dynamics is solved using the numerical exact {\it quasi adiabatic propagator path integral} (QUAPI)~\cite{QUAPI1,QUAPI2,QUAPI3,QUAPI4,QUAPI5}. At weak coupling we find damped oscillatory behavior as expected from RESPET. However, for biomolecular environments at temperature $T$ with $\Delta_0\simeq\omega_c\simeq T$ the numerically observed damping rate differs from the one-phonon rate of the analytical calculation by up to one order of magnitude. This difference increases with decreasing cut-off frequency $\omega_c$. This discrepancy cannot simply be addressed to multi-phonon processes since the damping rate behaves strictly linear in the coupling and multi-phonon processes come only into play at stronger couplings. We conclude in section~\ref{sec5} accordingly that a weak coupling Markovian approximation qualitatively describes the dynamics but fails quantitatively by up to an order of magnitude. This failure is nonperturbative in nature and can thus not be corrected by higher order perturbative approaches. Strictly spoken it invalidates all approaches to discuss the exciton dynamics in FMO which base on perturbative arguments. It even raises question marks on nonperturbative approaches which give at weak coupling comparable results to Redfield approaches.

\section{Model for biomolecular energy transfer}\label{sec2}

The simplest way to model a single chromophore (or pigment) is by describing it as a two-level system consisting of a ground and an excited state. When the electronic ground state is excited the electron is not free but localized by its attractive interaction with the hole it left. This dipole electron-hole configuration forms an exciton. A formal description can be given as
\be H \,=\, E\frac{\tau_z}{2} \ee
where we introduced the Pauli matrix $\tau_z$. Environmental fluctuations will cause transitions between the ground and the excited state and will add a fluctuating energy. Experimentally it is known that the recombination time is of the order of nanoseconds whereas the complete energy transfer through the complex is of the order of picoseconds. Thus the environmental fluctuations causing recombination are negligible. Describing the fluctuations by harmonic oscillators, which couple linearly to the chromophore, results in the independent boson model for a single chromophore~\cite{Gilmore05,Gilmore06,Gilmore08}
\be H \,=\, \epsilon\frac{\tau_z}{2} \,-\frac{\tau_z}{2}\sum_k\nu_kq_{k} \,+\halb\sum_k \left(p_{k}^2+\omega_k^2q_{k}^2 \right)
\ee
where we introduced the position and momentum operators, $q_k$ and $p_k$, of the mode with wave vector $k$ and its coupling $\nu_k$ to the chromophore and fixed $\hbar=\kb=1$ which we keep below.

We are not aiming at a description of the exciton transfer dynamics in the FMO complex but are interested in a thorough discussion of the validity of weak coupling perturbative approaches. 
For the sake of simplicity we thus restrict the model under consideration to just two chromophores. Since recombination is irrelevant for the transfer process of a single exciton we can restrict our investigation to two states $|i\rangle$ with the exciton at chromophore $i$. Furthermore neglecting the differing site energies leads to a donor-acceptor Hamiltonian~\cite{Gilmore05,Gilmore06,Gilmore08}
\be
H \,=\, \frac{\Delta}{2}\sigma_{x} \,-\frac{\sigma_z}{2}\sum_k\lambda_kq_k \,+\halb\sum_k \left(p_k^2+\omega_k^2q_k^2 \right)
\ee
with tunneling element $\Delta$ between donor and acceptor and 
by introducing the Pauli matrices $\{\bbbone,\sigma_x,\sigma_y,\sigma_z\}$ with $\sigma_x=|1\rangle\langle 2| +|2\rangle\langle 1|$ and $\sigma_z=|1\rangle\langle 1| -|2\rangle\langle 2|$ and 
\be \lambda_k \,:=\, \nu_k({\bf r}_1)-\nu_k({\bf r}_2)\, .
\ee
Assuming a distance $|r_1-r_2|>hc/\Delta$ larger than the wave length of resonant modes with $c$ being speed of sound the position dependence is negligible and the spectrum
\be J(\omega) \,=\, \sum_k\frac{\lambda_k^2}{2\omega_k} \delta(\omega_k-\omega) \,=\, 2 \sum_k\frac{|\nu_k|^2}{2\omega_k} \delta(\omega_k-\omega) \, .
\ee
The paramount approach to discuss the dynamics of a donor-acceptor model is the NIBA~\cite{SpiBoLe1987,Weiss99}. It specifically succeeds to describe the {\it scaling} limes, $\omega_c\rightarrow\infty$, but is less trustable for bioenvironments with $\Delta\simeq\omega_c$. Besides numerical approaches\cite{BioTh2008} the only other analytical rigorous approach is a weak coupling perturbative treatment which provides a treatment formally not limited to the {\it scaling} limes.

We focus on the occupation difference between donor and acceptor $P(t)=\langle\sigma_z\rangle(t)$ with initially the exciton on the donor $P(0)=1$~\cite{foot3}. Without coupling to the bath the occupation difference shows oscillations, $P(t)=\cos(\Delta t)$ with frequency $\Delta$. The bath primarily causes damping of the oscillations. In order to determine the bath influence we calculate the time evolution operator $\U(t)$ with
\be \U(t) \rho_0 \,=\, \rho(t) \ee
with the statistical operator $\rho(t)$ obeying
\be\label{neumann} \partial_t \rho(t) \,=\, -i[H,\rho(t)] \,=:\, \L \rho(t) \ee
with initial condition $\rho_0$ and Liouville operator $\L$ in lowest relevant order in the coupling between spin and bath. We show in the next section that two additional weak contributions emerge whose residua are proportional to $\alpha$ and which are negligible under usual experimental conditions.

\section{Analytical approach for weak system-bath coupling - RESPET}\label{sec3}

For a perturbative approach in the coupling between spin and bath we split the Hamiltonian
\be H_0 \,=\, H \,-H_{SB} \with H_{SB} \,=\, -\frac{\sigma_z}{2}\sum_k\lambda_kq_k \ee
and likewise the Liouvillian $\L=\L_0+\L_{SB}$. The time evolution operator obeys a Dyson equation
\bee \U(t,t_0) &=& \U_0(t,t_0) \,+\int_{t_0}^t ds U_0(t,t_0) \L_{SB} \U_0(s,t_0) \nonumber\\
&&\hspace*{-1.5cm} +\int_{t_0}^t ds\int_{t_0}^s ds' U_0(t,t_0) \L_{SB} \U_0(s,s')\L_{SB} \U(s',t_0)
\eee
with $\U_0(t,t_0)$ the time evolution of the uncoupled spin and bath. We are interested in the dynamics of the spin alone allowing us to integrate out the bath degrees of freedom in the Dyson equation resulting in
\bee\label{Ueff} \U_{\rm eff}(t,t_0) &=& \U_S(t,t_0) \nonumber\\
&&\hspace*{-1.0cm}+\int_{t_0}^t ds\int_{t_0}^s ds' U_S(t,t_0) \M(s,s') \U_{\rm eff}(s',t_0)
\eee
with the memory kernel $M(s,s')=\langle\L_{SB} \U_0(s,s')\L_{SB}\rangle_B$ where $\langle \cdot\rangle_B=\spur{\cdot \rho_B}$ with the initial bath statistical operator $\rho_B$ assuming a factorizable initial condition $\rho_0=\rho_S\otimes\rho_B$. Herein it is essential that the linear order term $\langle \L_{SB}\rangle_B=0$ vanishes. The memory kernel is given in lowest order (1-loop approximation) in the coupling between system and bath only~\cite{foot1}.

With the Laplace transformation defined as
\be f(z)\,=\, i\int_0^\infty dt e^{izt}f(t) \ee
and
\be f(t)\,=\, \frac{1}{2\pi i}\int_{-\infty}^\infty dz e^{-izt}f(z)
\ee
the Dyson equation translates into
\be \U_{\rm eff}(z) \,=\, \left( \U^{-1}_S(z) \,+ \M(z) \right)^{-1}
\ee

For a single two level system or spin the space for the time evolution operator $\U_{\rm eff}$ is spanned by basis $\{\bbbone, \sigma_x, \sigma_y, \sigma_z\}$ and the operators $\U_{\rm eff}$ as well as $\U_S$ and $\M$ can be expressed as 4$\otimes$4 matrices
\be \U_S(z) \,=\, \left( \begin{array}{c|c} \begin{array}{cc} -z^{-1} & 0 \\ 0 & -z^{-1} \end{array} & 0 \\\hline
0 & \begin{array}{cc} \frac{-z}{z^2-\Delta^2} & \frac{i\Delta}{z^2-\Delta^2} \\ \frac{-i\Delta}{z^2-\Delta^2} & \frac{-z}{z^2-\Delta^2} \end{array} \end{array} \right)
\ee
\be \M(z) \,=\, \left( \begin{array}{c|c} \begin{array}{cc} 0 & 0 \\ -\phi(z) & -\psi(z) \end{array} & 0 \\\hline
0 & \begin{array}{cc} -2\Gamma(z) & 0 \\ 0 & 0 \end{array} \end{array} \right)
\ee
with 
\bee \phi(z) &=& B_R(z-\Delta)-B_R(z+\Delta) \nonumber\\
\psi(z) &=& 2\{B_C(z-\Delta)+B_C(z+\Delta)\} \nonumber\\
\Gamma(z) &=& 2B_C(z) \nonumber
\eee
and the bath correlation function
\be\label{BC} B_C(z) \,=\, \viertel \int_0^\infty d\omega J(\omega)\coth(\beta\omega/2)\frac{-z}{z^2-\omega^2} \ee
and the bath response function
\be\label{BR} B_R(z) \,=\, \viertel \int_0^\infty d\omega 2J(\omega)\frac{-\omega}{z^2-\omega^2} \ee
with $J(\omega)=\sum_k(\lambda_k^2/2m\omega_k)\delta(\omega-\omega_k)=2\alpha\omega\exp(-\omega/\omega_c)$.

Finally we obtain the effective time evolution operator
\be \U_{\rm eff}(z) \,=\, \left( \begin{array}{c|c} \begin{array}{cc} -z^{-1} & 0 \\ a(z) & b(z) \end{array} & 0 \\\hline
0 & \begin{array}{cc} \frac{z}{p(z)} & \frac{-i\Delta}{p(z)} \\ \frac{i\Delta}{p(z)} & \frac{z+2\Gamma(z)}{p(z)} \end{array} \end{array} \right)
\ee
with 
\bee a(z) &=& -(\phi(z)/\psi(z))\{(z+\psi(z))^{-1} -z^{-1} \} \nonumber\\
b(z) &=& -(z+\psi(z))^{-1}\nonumber
\eee
and
\be p(z) \,=\, \Gamma^2(z)-\{ z+\Delta+\Gamma(z)\}\{ z-\Delta+\Gamma(z)\} \ee

We focus on the occupation difference of the two level system
\be P(t) \,:=\, \langle \sigma_z(t)\rangle_S \ee
where $\langle \cdot\rangle_S=\spur{\cdot \rho_S}$ with the systems initial statistical operator $\rho_S=\halb(\bbbone+\sigma_z)$ defining $P(0)=1$ as initial condition. With the definition of correlation operators $\bar{\sigma_i}$ by $\bar{\sigma_i}\sigma_j=\halb\{\sigma_i\sigma_j+\sigma_j\sigma_i\}$ one can express the occupation difference by
\be P(z) \,=\, \langle \bar{\sigma}_z\U_{\rm eff}(z)\rangle_S \,=\, \frac{z+2\Gamma(z)}{p(z)} \ee

Thus we have formally solved the dynamics of the two level system. What is left is the Laplace back transformation:
\be P(t) \,=\, \frac{1}{2\pi i}\int_{-\infty}^\infty dz e^{-izt} \frac{z+2\Gamma(z)}{p(z)} \;. \ee
The integration is closed in the lower half plane and thus we need to know the branch cuts and poles of $P(z)$ in the lower half plane. Keep in mind that the Laplace transforms are analytical in the upper half plane by definition. 

The integral representation of the bath correlation (\ref{BC}) is not valid in the lower half plane since it has a branch cut along the real axis. In order to continue the function into the lower plane we split it according to $B_C=B_C^++B_C^-$ with
\be B_C^\pm(z) \,=\, \achtel \int_0^\infty d\omega J(\omega)\coth(\beta\omega/2)\frac{-1}{z\mp\omega} \ee
where $B_C^\pm$ has its branch cut only from $0$ to $\pm\infty$ and thus can trivially be continued for $Re(z)\lessgtr 0$. For $Re(z)>0$ we need to shift the path of integration in the integral representation of $B_C^+$ in order to continue the function into the lower half plane leading to
\bee B_C^\pm(z) &=& \achtel \int_0^\infty d\omega J(\omega)\coth(\beta\omega/2)\frac{-1}{z\mp\omega} \nonumber\\
&&\hspace*{-0.8cm}+ i\viertel\pi J(\pm z)\coth(\beta (\pm z)/2) \Theta\left({\rm Im}(z)\lessgtr 0\right) \eee
with the Heavyside function $\Theta(\cdot)$.
Since the integrand, i.e. the hyperbolic cotangent, has poles on the negative imaginary axis the path of the integrand can maximally be shifted to run along the negative imaginary axis. Thus the bath correlation function can be presented having a branch cut along the negative imaginary axis and as well poles on it due to the term $\coth(\beta (\pm z)/2)$.

\subsection{System poles}

With this prerequisite we can tackle the Laplace back transformation. We expect generally three contributions. First without coupling to the bath the dynamics is governed by the two system poles $z_\pm=\pm\Delta$ leading to $P(t)=\cos(\Delta t)$. At weak coupling we expect the main behavior still governed by these two poles but we further expect corrections to the frequency, a decay rate and the prefactor diminishing this contribution to the total dynamics since the branch cut as well as the poles of the bath correlation function will result in additional contribution.

Without going into details the system pole contribution becomes
\bee\label{syspol} P^S(t) &=& \left( 1+\partial_z {\rm Im}(\Gamma(\Delta)) +\frac{{\rm Re}(\Gamma(\Delta))}{\Delta}\right) \, \cos(\tilde{\Delta}t) e^{-\Gamma_0t} \nonumber\\
&& +\left(\partial_z {\rm Re}(\Gamma(\Delta))+\frac{{\rm Im}(\Gamma(\Delta))}{\Delta}\right) \, \sin(\tilde{\Delta}t) e^{-\Gamma_0t}
\eee
with $\tilde{\Delta}=\Delta+{\rm Re}(\Gamma(\Delta))$ and $\Gamma_0={\rm Im}(\Gamma(\Delta))=\viertel\pi J(\Delta)\coth(\beta\Delta/2)$. A standard Markov approximation stops at this point since it focuses on the dynamics resulting from the system poles. Within the presented weak coupling approach the other contributions can also be determined.

\subsection{Bath poles}

The second contribution comes from the poles of the bath correlation function since around them $\Gamma(z)$ will become very large and accordingly
\be p(z) \,=\, \Delta^2-z^2-2 \Gamma(z)z \ee
will have zeros. The poles are at $-it_n=-i2\pi n T$ with $n>0$ an integer. Focusing first on ${\rm Re}(z)>0$ we can safely assume that the integral part in the continuated bath correlation function will be small at the pole of the second part and thus we obtain
\be \Gamma(z=-it_n+\epsilon_n^+) \,\simeq\, \halb\pi iJ({\rm Im}(\Gamma(\Delta))) \frac{2 T}{\epsilon_n^+}
\ee
which leads us to zeros of $p(z_n)$ at $z_n=-it_n+\epsilon_n^+$ with
\be \epsilon_n^+ \,=\, \frac{\pi t_n J(-it_n)2T}{t_n^2+\Delta^2} \ee
Depending on the ratio $t_n/\omega_c$ the real part of $\epsilon_n^+$ turns out to be positive or negative. Whenever it is negative for a certain $n$ this pole does not contribute since we restricted the investigation to ${\rm Re}(z)>0$. Similarly we get
\be \epsilon_n^- \,=\, -\frac{\pi t_n J(it_n)2T}{t_n^2+\Delta^2} \ee
which only results in a pole when ${\rm Re}(\epsilon_n^-)<0$. These conditions are only fulfilled for both identically ensuring that all poles come in pairs resulting in a real time-dependent function. The residua of these contributions can be estimated by
\be \mbox{Res}\left\{ \frac{z+2\Gamma(z)}{p(z)} \right\} \,=\, {\rm Re}\left(\frac{\Delta^2 T}{(t_n^2+\Delta^2)^2}\, i\viertel J(-it_n) \right) \,=:\, \nu_n
\ee
finally leading us to
\be P^B(t) \,=\, \sum_{n=1}^\infty\nu_n \cos({\rm Re}(\epsilon_n^+)t) e^{-(t_n+{\rm Im}(\epsilon_n^+)t}
\ee
With increasing $n$ the weights $\nu_n$ decrease and the relaxation times increase for these terms. In the case $T\sim \Delta\sim \omega_c$, we are interested in, only the very first few $n$'s are of any relevance. The relaxation time is of the same order as the tunneling element $\Delta$ and thus this term is fast decaying and weak due to $\nu_n$ to begin with. Its frequency, however, is small too, thus when $T \simeq \Delta$ on a time scale of $\Delta^{-1}$ this contribution will be present as an overdamped behavior for not too weak couplings. This contribution is clearly a bath generated dynamics which is non-Markovian in nature.
\begin{figure}[t]
\epsfig{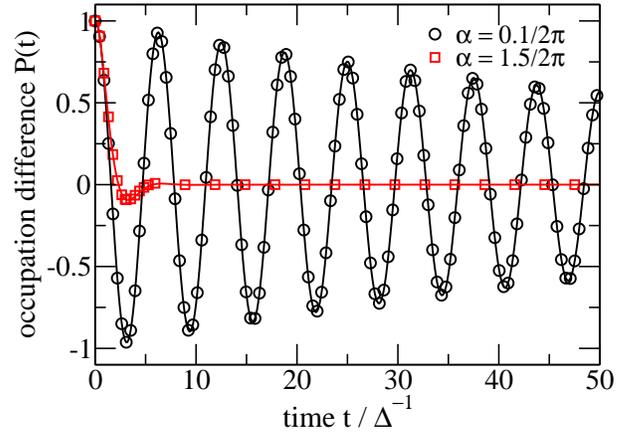}
\caption{\label{fig0} The occupation difference $P(t)$ versus time for $T=\omega_c=1.5 \Delta$. Symbols represent data (most points not printed) obtained by QUAPI and the full lines are fits to the data.}
\end{figure}

\subsection{Branch cut}

The third contribution to the occupation difference $P(t)$ stems from the branch cut of the bath correlation function which generates a similar branch cut along the negative imaginary axis for $P(z)$. Firstly we have to calculate the jump at the branch cut
\be S(z=-is) \,=\, \lim_{\eta\rightarrow 0} \{ P(\eta-is) - P(-\eta-is) \}
\ee
Close to the poles $\Gamma(z)$ dominates and the jump vanishes. As long as we are far from the poles of $\Gamma(z)$ we can approximate
\be S(z=-is) \,\simeq\, -2\pi \frac{{\rm Re}\{J(-is)\}\cot(\beta s/2)}{s^2+\Delta^2} \ee
and then obtain
\be P^C(t) \,=\, \frac{-1}{2\pi}\int_0^\infty ds e^{-st} S(-is)
\ee
We are interested in times $t\gtrsim\Delta^{-1}$ and $T \sim\Delta \sim \omega_c$ and thus the exponential in the integral allows us to restrict ourselves to $s\lesssim\Delta$ and to approximate $S(-is) \,\simeq\, -4\pi \alpha s (2T/\Delta^2/\omega_c)$ and thus
\be P^C(t) \,=\, 4\alpha \frac{T}{\omega_c}\, \frac{1}{(\Delta t)^2}
\ee
This contribution could actually be rather large when $(T/\omega_c)>1$ and its decay is algebraic and thus slow compared to the exponential decay of the other terms. However, we are restricted to times $t\gg\Delta^{-1}$ rendering its contribution small.
\begin{figure}[t]
\epsfig{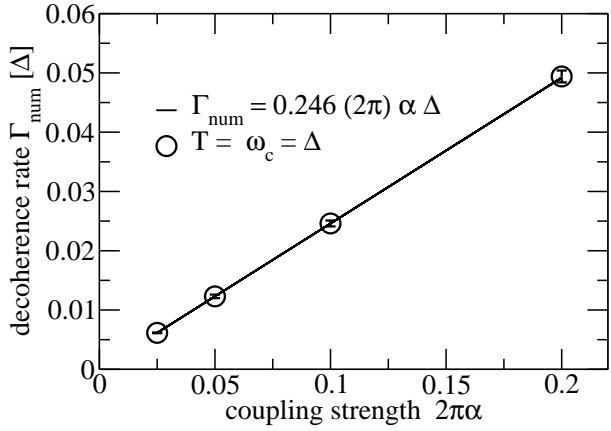}
\caption{\label{fig1} The decoherence rate $\Gamma_{\rm num}$ versus coupling constant $\alpha$ at $T=\Delta=\omega_c$ including error bars for the numerical procedure. The line is a linear fit to the data.}
\end{figure}

All in all, we have three contributions to the dynamics of the two level system. We denote them in the following as the system pole, the bath pole and the bath branch cut dynamics. The contribution of the later two is proportional to the coupling constant $\alpha$ and thus small in weak coupling. These are the non-Markovian parts of the dynamic. An estimate of the three parts for $\Delta=T=\omega_c$ at a time $t_1=2\Delta^{-1}$ is obtained as
\bee P^C(t_1) &=& \alpha \nonumber \\
P^B(t_1) &=& \sum_{n=1}^\infty\frac{\pi n \alpha e^{-4\pi n}}{2(4\pi^2n^2+1)}  \,\simeq\, \frac{\pi \alpha e^{-4\pi} }{2(4\pi^2+1)} \nonumber \\
P^S(t_1) &\le & \exp(-\pi\alpha e^{-1}\coth(\halb)) \,\simeq\, e^{-2.5\alpha} \nonumber
\eee
and shows that the non Markovian parts are only small corrections at weak coupling. Thus, one would conclude that at weak coupling Markovian dynamics and a one-phonon process dominate. In the next section we determine the occupation difference using the numerical exact QUAPI~\cite{QUAPI1,QUAPI2,QUAPI3,QUAPI4,QUAPI5} approach in order to test the validity range of the weak coupling approach.

\section{Numerical approach: exact results}\label{sec4}

Using the {\it quasi adiabatic path integral} (QUAPI)~\cite{QUAPI1,QUAPI2,QUAPI3,QUAPI4,QUAPI5}, a numerically exact approach, we determined the time dependent occupation difference $P(t)=\langle\sigma_z\rangle(t)$ of the donor-acceptor system coupled to environmental fluctuations with Ohmic spectrum and $\Delta\sim\kb T\sim\omega_c$. 
The occupation difference $P(t)$ shows weakly damped oscillations which we fitted with
\be P(t) \,=\, \left\{ \cos(\Delta_{\rm num}t) + \mu \sin(\Delta_{\rm num}t)\right\} e^{-\Gamma_{\rm num} t} 
\ee
using the tunneling frequency $\Delta_{\rm num}$, the damping or decoherence rate $\Gamma_{\rm num}$ and a prefactor $\mu$ as fitting parameters. 
\begin{figure}[t]
\epsfig{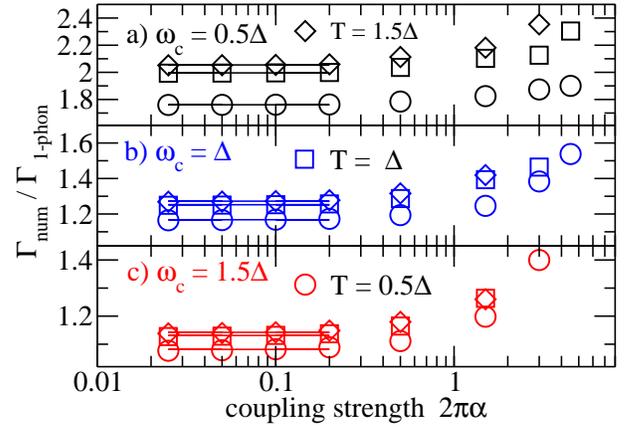}
\caption{\label{fig2} The ratio $\Gamma_{\rm num}/\Gamma_{\rm 1-phon}$ versus coupling constant $\alpha$}
\end{figure}

According to our weak coupling analytical result the prefactor $\mu$ is proportional to the coupling $\alpha$. The fitting accuracy, however, is rather insensitive to $\mu$. Besides it is small for small $\alpha$ and for large $\alpha$ the occupation difference is dominated by the exponential decay. Therefore we fixed $\mu=\Gamma_{\rm num}/\Delta_{\rm num}$ as known for large $\omega_c$~\cite{SpiBoLe1987,Weiss99}. The modification of the tunneling frequency, 
$|\Delta_{\rm num}-\Delta|/\Delta$, due to the coupling to the environmental fluctuations is less than 5\% in the whole studied parameter range. This illustrates that for $\omega_c\lesssim\Delta$ the effect of the fluctuations on $\Delta$ cannot be included by a mere parameter renormalization $\Delta\rightarrow\Delta_{\rm eff}$~\cite{SpiBoLe1987,Weiss99}, as it is the case in the regime $\omega_c\gg\Delta$. The main relevant effect of the environment is the emergence of a decoherence rate which we discuss in the following.

The chosen fit function was sufficient to describe all data discussed. Fig.~\ref{fig0} shows exemplary two data sets with $T=\omega_c=1.5 \Delta$ for a weak coupling $\alpha=0.1/(2\pi)$ and a medium coupling $\alpha=1.5/(2\pi)$ with the respective fits.

Now we focus solely on a discussion of the decoherence rate. Fig.~\ref{fig1} plots the decoherence rate at temperature $T=\Delta$ and for a cut-off frequency $\omega_c=\Delta$ versus coupling $\alpha$. The line in fig.~\ref{fig1} is a fit to the numerical data giving clear evidence for the linear dependence of the decoherence rate on coupling strength: $\Gamma_{\rm num}=0.246(2\pi)\alpha\Delta$. Our weak coupling result predicted $\Gamma_0=\viertel\exp(-\Delta/\omega_c)\coth(\Delta/2\kb T)2\pi\alpha\Delta\simeq 0.199 (2\pi)\alpha\Delta$. Thus the numerically extracted decoherence rate is about 25\% larger than the weak coupling expression predicts for {\it all} investigated coupling strengths. 

At first glance one might attribute these differences to multi-phonon processes which are beyond the analytical one-phonon rate. However, multi-phonon processes generally are not linear in the coupling $\alpha$ and accordingly the decoherence rate should not be linear in $\alpha$. In order to see the onset of multi-phonon processes at higher couplings Fig.~\ref{fig2} shows the ratio of the numerically observed decoherence rate to the 1-phonon rate versus the coupling strength $\alpha$ for three temperatures and cut-off frequencies. At weak coupling, $\alpha\le0.2/(2\pi)$ the ratio is constant. At medium couplings $0.2/(2\pi)\le\alpha\le 0.5$ the ratio slightly increases with coupling showing the onset of multi phonon processes (higher order processes in $\alpha$)~\cite{foot2}. Most importantly, the deviations of the ratio from $1$ are independent of the coupling strength for weak couplings. The deviations increase slightly with increasing temperature and rather strongly with decreasing cut-off frequency $\omega_c$.

Fig.~\ref{fig1} and Fig.~\ref{fig2} together show that in the parameter regime $\Delta\sim\kb T\sim\omega_c$ relevant for bioenvironments the weak coupling approach fails even at lowest coupling. Since the deviations between the 1-phonon rate and the numerical exact one is independent of the coupling $\alpha$ (for weak coupling) it is not a result of higher order processes and thus perturbatively not tractable. Accordingly all perturbative approaches (even in higher order) are bound to fail to describe the dynamics correctly beyond a mere qualitative description.

\begin{figure}[t]
\epsfig{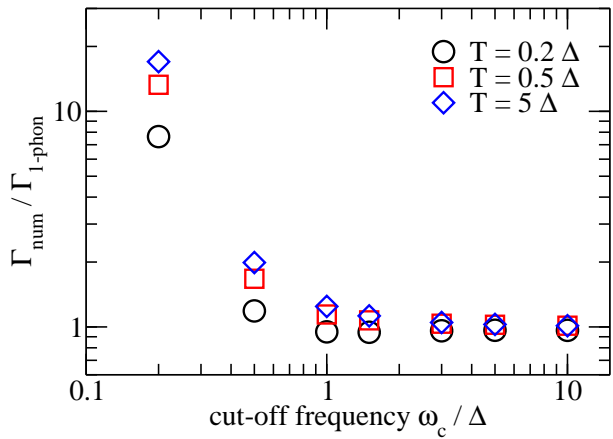}
\caption{\label{fig3} The ratio $\Gamma_{\rm num}/\Gamma_{\rm 1-phon}$ versus cut-off frequency $\omega_c$ for three different temperatures for a fixed weak coupling $\alpha=0.1/(2\pi)$.}
\end{figure}

In order to show the full extent of this failure Fig.~\ref{fig3} plots the ratio $\Gamma_{\rm num}/\Gamma_{\rm 1-phon}$ versus cut-off frequency $\omega_c$ for three different temperatures for a fixed weak coupling $\alpha=0.1/(2\pi)$. For large cut-off frequencies $\omega_c\gg\Delta$ the ratio tends to $1$ and the one-phonon process accurately describes decohering dynamics. With decreasing cut-off frequency the ratio increases strongly and at $\omega_c=\Delta/5$ the numerical exact rate is an order of magnitude larger than the one-phonon process predicts.

Beyond our RESPET result we compare the numerical results with, one might be tempted to use available more sophisticated analytical treatments~\cite{SpiBoLe1987,Weiss99}. These  give in the limit of weak coupling expressions for the decoherence rate $\Gamma_0$ formally identical to our result (and thus Fermis golden rule) but the tunneling element $\Delta$ is exchanged by a renormalized one, $\Delta_{\rm eff}\propto (\Delta/\omega_c)^{\alpha/(1-\alpha)}$, depending on the cut-off frequency. At first we should remark that this renormalization decreases the decoherence rate for decreasing $\omega_c$ (in comparisson to $\Gamma_{\rm 1-phon}$) which is in clear contrast to our finding and secondly the adiabatic renormalization leading to the expression of $\Delta_{\rm eff}$ is justified only in the limit $\Delta\ll\omega_c$ whereas our observed deviations occur in the opposite limit.

\section{Conclusion}\label{sec5}

Exciton transport in photosynthesis is strongly disturbed by the fluctuations of the dipolar solvent for the biomolecules building the light-harvesting complexes. However, the fact that these fluctuations are slow, $\omega_c\simeq\Delta$, results in far longer decoherence times than known for solid state excitons where $\omega_c\gg\Delta$ and finally allows for enhanced quantum coherence in such systems~\cite{BioTh2008}. To study the extent and relevance of quantum coherence on exciton transport, however, systems like the FMO complex must be studied beyond a donor-acceptor model. Numerically exact treatments, like QUAPI or QMC, become rather time consuming when including all 7 chromophores and one is restricted to analytical approximations. Slow but strongly coupled environmental fluctuations cause strongly non-Markovian dynamics and thus approximative analytical approaches must be able to include such effects. To verify such methods they are typically matched to weak coupling perturbative results and to F\"orster transfer dynamics in the weak tunneling limit~\cite{Flemming2}.

We have determined the population difference in a donor-acceptor model in lowest order in the coupling between donor-acceptor and environment including all non-Markovian contributions and compare it to results from a numerical exact treatment using QUAPI. We show that in the experimentally relevant parameter regime, $\Delta\sim\kb T\sim\omega_c$, where there is {\it no small parameter}, the decoherence rate strongly differs from the one-phonon rate and the deviation increases with decreasing cut-off frequency $\omega_c$. This was shown for a weak coupling where the decoherence rate varies strictly linear with the coupling. Multi-phonon contributions showing non-linear dependence on the coupling are found for larger couplings as well. The deviations also increase with increasing temperature.

Thus we conclude that analytical approximations to discuss the influence of bioenvironmental slow fluctuations on quantum systems cannot be verified in the weak coupling limit by comparing to weak coupling perturbative results. 

The increasing deviations with decreasing cut-off frequency $\omega_c$ points to strong nonperturbative dynamics for small cut-off frequencies. We interpret this behavior as the breakdown of a picture where the slow environment acts as a bath which only exchanges energy with the quantum system. The fact, however, that the dynamics can still be described by damped coherent oscillations with a decoherence rate $\Gamma_{\rm num}$ varying linear with coupling constant $\alpha$ shows that at weak coupling a slow bath {\it renormalizes} itself to an effective bath whose influence can be described by an effective {\it one-phonon} process with renormalized bath parameters. An analytical description of this {\it renormalized} bath would be needed to base approximations on which allow to study the exciton transport dynamics in photosynthetic light-harvesting complexes.

We gratefully acknowledge support by the Excellence Initiative of the German Federal and State Governments.


\begin{thebibliography}{99}
\bibitem{foot4} Perturbative does not refer to time-dependent Hamiltonian perturbation theory but to a perturbative evaluation of the memory kernel as defined in Eq.~(\ref{Ueff}).
\bibitem{foot5} The exponential cut-off function is choosen for convenience. Specific high-frequency parts of the spectra change results quantitatively but not qualitatively.
\bibitem{foot3} Within linear order in the coupling strength $\alpha$ between system and bath $P(t)$ is also identical to the symmetrized correlation function $C_{zz}(t):=\langle \halb\{\sigma_z(t)\sigma_z(0)+\sigma_z(0)\sigma_z(t)\} \rangle_S$ with $\langle \cdot\rangle_S=\spur{\cdot \rho_S}$ and the systems initial statistical operator $\rho_S$ assumed to be in thermal equilibrium.
\bibitem{foot1} A possible way to prove equation (\ref{Ueff}) is to use the infinite Dyson series, then averaging and rearranging terms using Wicks theorem.
\bibitem{foot2} With increasing temperature and cut-off freqeuncy the dynamics at higher coupling tends towards overdamped behaviour where our numerical investigation does not meet QUAPI convergence criteria anymore. Only converged results are shown.
\end{thebibliography}
\end{document}